 \documentclass[preprint2]{aastex}

\newcommand{\lae}{\mathrel{<\kern-1.0em\lower0.9ex\hbox{$\sim$}}}
\newcommand{\gae}{\mathrel{>\kern-1.0em\lower0.9ex\hbox{$\sim$}}}
\shorttitle{{\it INTEGRAL} and {\it XMM} Spectral Studies of NGC 4388}
\shortauthors{Beckmann et al.}

\begin{document}

\title{{\it INTEGRAL} and {\it XMM-Newton} Spectral Studies of NGC 4388}


\author{V. Beckmann\altaffilmark{1}, N. Gehrels}
\affil{NASA Goddard Space Flight Center, Code 661, Greenbelt, MD 20771, USA}
\email{beckmann@milkyway.gsfc.nasa.gov}
\author{ P. Favre\altaffilmark{2}, R. Walter\altaffilmark{2}, T. J.-L. Courvoisier\altaffilmark{2}}
\affil{INTEGRAL Science Data Centre, Chemin d' \'Ecogia 16, 1290
 Versoix, Switzerland}
\author{P.-O. Petrucci}
\affil{Laboratoire d'Astrophysique de Grenoble, BP 53X, 38041 Grenoble Cedex, France}
\and
\author{J. Malzac\altaffilmark{3}}   
\affil{Centre d'\'Etude Spatiale des Rayonnements, 31028 Toulouse, France}


\altaffiltext{1}{Joint Center for Astrophysics, Department of Physics, University of Maryland, Baltimore County, MD 21250, USA}
\altaffiltext{2}{Observatoire de Gen\`eve, 51 Ch. des Maillettes, 1290 Sauverny, Switzerland}
\altaffiltext{3}{Institute of Astronomy, University of Cambridge, Madingley Road, Cambridge CB3 0HA, United Kingdom}

\begin{abstract}
We present first {\it INTEGRAL} and {\it XMM-Newton} observations of a Seyfert galaxy, the type 2 AGN NGC 4388. Several {\it INTEGRAL} observations performed in 2003 allow us to study the spectrum in the 20 - 300 keV range. In addition two {\it XMM-Newton} observations give detailed insight into the 0.2 - 10 keV emission. The measurements presented here and comparison with previous observations by {\it BeppoSAX}, {\it SIGMA} and {\it CGRO}/OSSE show that the overall spectrum from soft X-rays up to the gamma-rays
 can be described by a highly absorbed ($N_H \simeq 2.7 \times 10^{23} \, \rm cm^{-2}$) and variable non-thermal component in addition to constant non-absorbed 
thermal emission ($T \simeq 0.8 \, \rm keV$) of low abundance ($Z \sim 7 \% Z_\odot$), plus a constant Fe K$\alpha$ and K$\beta$ line. The hard X-ray component is well described by a simple power law with a mean photon index of $\Gamma = 1.7$. During the {\it INTEGRAL} observations the 20 -- 100 keV flux increased by a factor of 1.4. 
The analysis of {\it XMM-Newton} data implies that the emission below 3 keV is decoupled from the AGN and probably due to extended emission as seen in {\it Chandra} observations. The constant iron line emission is apparently also decoupled from the direct emission of the central engine and likely to be generated in the obscuring material, e.g. in the molecular torus.% or even further away.
\end{abstract}




\section{Introduction}
Seyfert 2 galaxies are among the optically faintest Active Galactic Nuclei (AGN) seen in the Universe. This is believed to be due to a geometrical effect. The emission of the super massive black hole at the AGN core is probably absorbed by a surrounding torus. The torus not only hides the broad line region, causing the characteristic Seyfert 2 optical spectra with narrow emission lines, but also absorbs most of the low energy X-ray emission. As the soft X-ray region is very sensitive to absorption in the line of sight, the study of the hard X-ray emission above $\sim 3 \rm \, keV$ is a powerful tool to investigate the central engine of Seyfert 2 (Ghisellini, Haardt \& Matt 1994).
In addition to this the variation of the Fe K$\alpha$ line seems to vary differently with respect to the continuum in several cases \citep{CHANDRANGC,fabian77,malzac}, which might be caused by the interaction of the core emission with the absorbing and reflecting material.

At higher energies absorption becomes less important revealing the emission from the close environment of the central black hole. This energy range could be dominated by comptonized hard X-ray radiation from a hot ($\sim 100 \, \rm keV$) plasma presumably forming an accretion disk corona \cite{diskcorona}, and Compton reflection of these hard X-rays on a cool accretion disk \citep{george,pexrav}. In Seyfert 2 galaxies, reflection of the central radiation on the molecular torus often leads to an additional, possibly dominant, reflection component (Ghisellini, Haardt \& Matt 1994).

NGC 4388 ($z = 0.00842$; Phillips \& Malin 1982) is a good probe to test these scenarios, as it is 
one of the brightest Seyfert 2 galaxies at hard X-rays. 
The energy range up to $\sim 10 \, \rm keV$ has been studied with X-ray missions like {\it ROSAT} \cite{ROSATNGC}, {\it ASCA} \citep{ASCAold,ASCANGC}, {\it BeppoSAX} \cite{SAXNGC}, {\it RXTE} (Sazonov \& Revnivtsev 2004), and lately with {\it Chandra} (Iwasawa et al. 2003). Spectral studies of NGC 4388 of the hardest X-ray energies up to several hundreds of keV have been performed by {\it SIGMA} \cite{SIGMA}, by the OSSE experiment (Johnson et al. 1993) on the {\it Compton Gamma Ray Observatory} (CGRO; Gehrels et al. 1993), and up to $\sim 150 \, \rm keV$ by the Phoswich Detector System (PDS; Frontera et al. 1997) on-board {\it BeppoSAX} \citep{sax,sax2}.
The two {\it BeppoSAX} observations in 1999 and 2000 showed a high energy spectrum with a power law of $\Gamma = 1.6$ and 1.5, respectively. Due to the energy range covered by the PDS it was only possible to give a lower limit for a possible high energy cut-off ($E_C > 109 \, \rm keV$).
The {\it INTEGRAL} mission \cite{INTEGRAL}, launched in October 2002, offers the unique opportunity to study the entire spectrum from 3 keV up to several MeV simultaneously. 
NGC 4388 was the first Seyfert 2 galaxy detected by {\it INTEGRAL} in January 2003 (Beckmann et al. 2004). In this paper we present data from several {\it INTEGRAL} observations and compare the results with previous {\it SIGMA}, OSSE, and {\it BeppoSAX} measurements. With a detection significance of $10.9 \sigma$ and $12.6 \sigma$ in the {\it INTEGRAL}/ISGRI energy bands $15 - 40 \rm \, keV$ and $40 - 100 \, \rm keV$, respectively, the data also allow us to study spectral variations between the 2003 observations. In addition {\it XMM-Newton} data are presented to study the connection with soft X-rays.

\section{Simultaneous X-ray and gamma-ray observations}
NGC 4388 was detected by {\it INTEGRAL} in the course of the 3C273 observation (Courvoisier et al.~2003a). Though NGC 4388 lies 10.4 degrees north of the quasar 3C273, valuable data for the Seyfert 2 galaxy were obtained due to the large field of view (FOV) of IBIS ($19^\circ \times 19^\circ$, partially coded FOV) and SPI ($35^\circ \times 35^\circ$, partially coded FOV), and because of the dithering observation strategy which is optimized for the spectrometer in order to allow a proper background determination. Observations were performed during seven {\it INTEGRAL} revolutions in January, June and July 2003 (Tab.~\ref{journal}). The total amount of exposure time is 512 ksec with 274 ksec and 238 ksec in the January and June/July campaign, respectively. Most of the observations were carried out in a dithering mode. 
36 ksec of observation were carried out in staring mode, and are therefore only partly useful for SPI analysis, but they allow a more precise flux extraction from the ISGRI data. Due to the smaller FOV of the two X-ray monitors (JEM-X) and of the optical monitor (OMC) on-board {\it INTEGRAL}, NGC 4388 has not been observed by those instruments. 
We will discuss the {\it INTEGRAL} data mainly in terms of the two observation campaigns in January 2003 (JAN03) and June/July 2003 (JUN03).
 
\subsection{IBIS}

The ISGRI spectrum has been extracted from ISGRI mosaics of the field
created in 20 energy bands. The count rate and statistical error were
extracted from the mosaic intensity and variance images for the ISGRI
source position derived from the highest significant pixel of the mosaic
in all energy bands.
%\footnote{\small http://lheawww.gsfc.nasa.gov/users/beckmann/spi/pages/ISGRI\_spectra.html}. 
The mosaic image fully redistributes the pixel values
of the sky images of individual science windows that were created using
version 3.0 ot the Offline Science Analysis (OSA) software distributed by the ISDC \cite{ISDC}. Pixel values are weighted according to vigneting and variance.
Finally we add a systematic error of 10\% to the spectra.

NGC 4388 was too faint to be seen by IBIS/PICsIT.
Summing all available IBIS/ISGRI data together shows that NGC 4388 is detected up to 200 keV, though the data above 150 keV are consistent with zero flux on a $1\sigma$ level. A fit to a single power law gives a photon index of $\Gamma = 1.70 {+0.01 \atop -0.01}$ (25 -- 300 keV) and a flux of $7.8 \pm 0.5 \, \rm mCrab$ in the 20 - 40 keV energy band.
Analysing the data according to the JAN03 and JUN03 campaign shows that the source was in a lower flux state in January 2003 with a harder spectral slope ($\Gamma = 1.61 {+0.19 \atop -0.18}$), while the June spectrum is steeper ($\Gamma = 1.76 {+0.16 \atop -0.14}$) but has a higher flux. Note however that the spectral shape is still consistent with the $\Gamma = 1.7$ model on a $1 \sigma$ level. For details see Table \ref{INTEGRALfit}.

\subsection{SPI}

In order to reduce the noise in the data reduction process of the SPI data, only those pointings were considered where NGC 4388 was less than 15 degrees off axis. This decreases the total amount of usable pointings to 122 with an exposure time of 178 ksec in JAN03 and 134 ksec in JUN03.
OSA 3.0 has been applied, except for some updates. In the binning of the data SPIHIST 3.1.2 was used in order to avoid a known problem with the earlier version.  In the case of the image reconstruction program SPIROS \cite{SPIROS}, we used the more recent version 6.0.1, and also a response function which takes into account the results from the in-flight calibration on the Crab (instrument response function spi\_irf\_rsp\_0015 and redistribution matrix spi\_rmf\_grp\_0002). For details of the SPI analysis procedures see Diehl et al. (2003). The SPI spectrum for the combined data of JAN03 and JUN03 is well described by a single power law with photon index $\Gamma = 1.68 {+0.47 \atop -0.35}$ (20 -- 500 keV) and a flux of $9 \pm 2.6 \, \rm mCrab$ in the $20 - 40 \, \rm keV$ energy band, which is consistent with the ISGRI results. 
Analysing the JAN03 and JUN03 subset it turned out that the source strength is not sufficient for SPI to achieve a reasonable signal to noise spectrum for the individual epochs.
In case of marginal detection the SPIROS software tends to overestimate the flux. This naturally leads also to a hardening of spectra, as the higher energy bins have a lower signal-to-noise. 
The results are shown in Table \ref{INTEGRALfit}. It can be clearly seen that while the flux value of the combined spectrum is similar to that of ISGRI, the subsets with the lower signal-to-noise show higher flux values, and this trend is stronger at higher energies.  
Although the trend in flux is similar to the one observed by ISGRI, the spectral evolution is reversed. But the large error bars on the SPI results have to be taken into account and the results are therefore still consistent with each other. For a detailed discussion of cross-calibration issues of the {\it INTEGRAL} instruments see Lubi\'nski et al. (2004).

\subsection{{\it RXTE}}
The All Sky Monitor (ASM) on board the {Rossi X-ray Timing Explorer} ({\it RXTE}) scans about 80\% of the sky every orbit. This offers an unique way to monitor the emission of bright X-ray sources like NGC 4388 in the $1.5 - 12 \rm \, keV$ energy range. We extracted fluxes from the {\it RXTE}/ASM data base. The fluxes have been averaged over the same time periods as the JAN03 and JUN03 {\it INTEGRAL} observations in order to have comparable results. 
%Only measurements for which the flux added to the $1\sigma$ error is larger than 0.0 have been taken into account, so that only reliable data are included. The weighted mean of the JAN03 data is $8.6 \pm 1.2 \, \rm mCrab$, while during the JUN03 campaign the flux was $6.7 \pm 0.8 \, \rm mCrab$.
The weighted mean of the JAN03 data was $3.5 \pm 1.2 \, \rm mCrab$, while during the JUN03 campaign the flux was $2.2 \pm 1.9 \, \rm mCrab$. Due to low statistics the analysis in the three different ASM bands did not give further information about the flux variability. 

\section{{\it XMM-Newton} observations}

Two observations with {\it XMM-Newton} (Jansen et al. 2001) were performed in July and December 2002 as described in Table \ref{journalXMM}. Concerning the connection between the soft and hard X-ray emission, the most interesting data for our study come from the two MOS cameras \cite{MOS} and from the PN detector \cite{PN}, as they cover the energy range $0.2 - 10 \rm \, keV$.
The data have been reduced using the {\it XMM-Newton} Science Analysis Software version 5.4.1 and the model fitting using XSPEC 11.3 was done simultaneously for the PN and MOS data. Both observations are well represented by a Raymond-Smith model \cite{raymond} at energies below $\sim 2.5 \, \rm keV$ with low abundance (7\% solar) and a temperature of about $0.80 \rm \, keV$, applying only Galactic absorption ($2.62 \times 10^{20} \, \rm cm^{-2}$). 
The so-called {\it mekal} model in XSPEC, developed by Mewe \& Kaastra, gives the same results ($T = 0.78 \, \rm keV$, $7.4 \% \, \rm Z_\odot$).

% $T = 0.779 {+0.044 \atop -0.044} \, \rm keV$, $7.4 {+2.0 \atop -1.7} \% \, \rm Z_\odot$
Around 2.5 keV 
another additional component is seen in the data. This can be modeled by allowing the abundances of the single elements involved to vary. It turns out that this part of the spectrum is best modeled when the sulfur abundance is about solar, thus being 20 times higher than the relative abundances of the other elements.

The data do not allow to verify the more complex model of a photoionized gas applied by Iwasawa et al. (2003) to the {\it Chandra} data. This model avoids the (unrealistic) low abundance, and therefore seems to be more appropriate from a physical point of view.

At harder X-rays, the spectrum is dominated by an absorbed power law. The column density of the cold absorber is $2.7 \times 10^{23} \, \rm cm^{-2}$. The power-law index is not well constrained by the {\it XMM-Newton} data, so the photon index of $\Gamma = 1.7$ measured at hard X-rays, has been applied. In addition, a gaussian line at $6.39 {+0.01 \atop -0.01} \rm \, keV$ is apparent in both observations. Details of the fit are listed in Table~\ref{XMMfit}. This is consistent with a recent measurement by {\it Chandra}, which showed an Fe line of centroid in 6.36 $+0.02 \atop -0.02$ keV \cite{CHANDRANGC}. Another weak line is detectable at $6.89  {+0.13 \atop -0.10} \rm \, keV$ (probably Fe K$\beta$) with a flux about ten times lower than the Fe K$\alpha$ line.
In both {\it XMM} spectra there was no sign of a further line at $7.1 \rm \, keV$, as reported from {\it Chandra} data by Iwasawa et al. (2003). However, due to the lower resolution of the {\it XMM} in this energy region, a contribution from a $7.1 \rm \, keV$ line might be included in the wing of the Fe K$\beta$ line.

\begin{figure}
\plotone{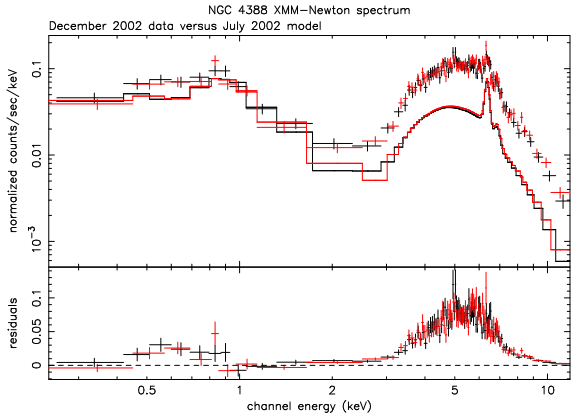}
\caption[]{Comparison of {\it XMM-Newton} MOS data from December 2002 with respect to the model for the July 2002 EPIC/MOS data.}
\label{fig:compareMOS}
\end{figure}

The variation between the July and December data occurs mainly in the high energy part. This is clearly seen when comparing the December data with the model for the July data (Fig.~\ref{fig:compareMOS}). While there are little or no changes below $\sim 2.5 \, \rm keV$, the spectrum varied significantly at higher energies. The iron fluorescence line at $6.39 \rm \, keV$ was not affected by the variations. 

Sazonov \& Revnivtsev (2004) studied the {\it RXTE}/PCA slew data of AGN in the $3 - 20 \, \rm keV$ energy band. This resulted for NGC 4388 in a luminosity of $L_{3-20 \, \rm keV} = 4.5 \times 10^{42} \, \rm erg \, s^{-1}$, which is inside the range of that observed by {\it XMM-Newton} in 2002 (extrapolated $L_{3-20 \, \rm keV} = (2.8 - 7.0) \times 10^{42} \, \rm erg \, s^{-1}$, for the July and December 2002 observation, respectively).

\section{X-ray to Gamma-ray spectrum}

The {\it INTEGRAL} data alone do not allow us to reconstruct the complete X-ray spectrum of NGC 4388, mainly because the statistics of the ISGRI and SPI data do not constrain existing models. 

\begin{figure}
\plotone{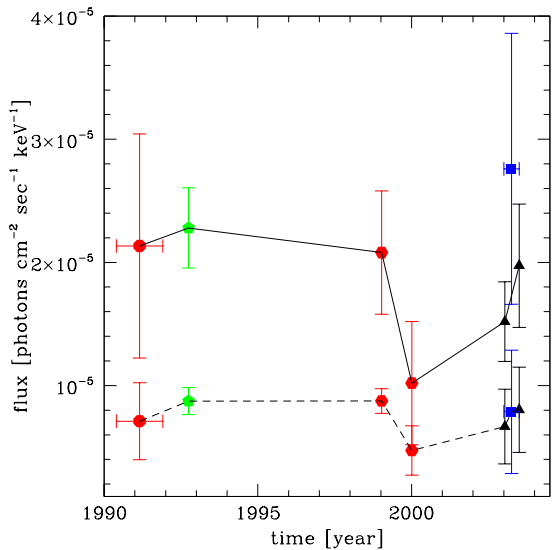}
\caption[]{Lightcurve for NGC 4388 at 60 keV (upper solid line) and at 100 keV (lower dashed line). Data taken from {\it SIGMA} (circle), {\it OSSE} (pentagon), {\it BeppoSAX}/PDS (hexagon), SPI (square), and ISGRI (triangle).}
\label{fig:lightcurve}
\end{figure}

Comparison with former observations by {\it SIGMA} \cite{SIGMANGC},  OSSE (Johnson et al. 1994), and {\it BeppoSAX}/PDS \cite{SAXNGC} shows a similar spectrum at hardest X-rays as observed with {\it INTEGRAL}. Measurements by BATSE gave a flux of $f_{20 - 100 \rm \, keV} = (2.6 \pm 1.5) \times 10^{-3} \, \rm ph \, cm^{-2} \, s^{-1}$, compared with an {\it INTEGRAL} value of $(2.7 \pm 0.3) \times 10^{-3} \, \rm ph \, cm^{-2} \, s^{-1}$. This indicates that the soft gamma-ray spectrum does not vary dramatically as shown in the lightcurves for the 60 keV and 100 keV flux in Fig.~\ref{fig:lightcurve}. The large error in the SPI flux at 60 keV is mainly based on the fact that a strong detector background line complex is apparent in the 53 -- 67 keV energy range resulting from $^{73}$Ge (Weidenspointner et al. 2003), which makes source flux extraction difficult. The flux varied by a factor $< 3$ and the spectral shape is conserved during the 13 years of various observations. 

In order to extend the spectrum into the energy range below 20 keV we added the two {\it XMM-Newton} EPIC data sets, described in the previous section. 

\begin{figure}
\plotone{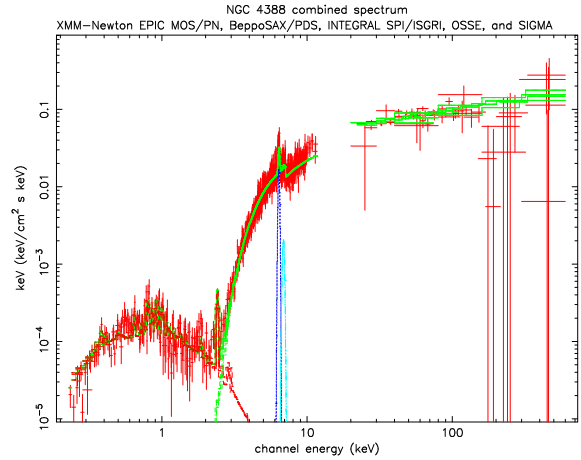}
\caption[]{NGC 4388 combined spectrum ($E^{2} f_E$ vs. $E$), including {\it XMM-Newton} EPIC MOS/PN, {\it BeppoSAX} PDS, {\it CGRO} OSSE, {\it SIGMA}, and {\it INTEGRAL} ISGRI/SPI. The high energies are dominated by the highly absorbed reflection component plus the K$\alpha$ and K$\beta$ iron fluorescence lines, while the lower energies show emission characteristic for a hot plasma with low abundance. XMM data from July 2002 are not shown in this plot for better visibility.}
\label{fig:combined}
\end{figure}

The data were fit simultaneously in XSPEC 11.3 applying the same model as for the {\it XMM} data alone: a Raymond-Smith model for a hot plasma with Galactic absorption, an absorbed power law, 
plus two Gaussian emission lines, representing the K$\alpha$ and K$\beta$ line. Figure~\ref{fig:combined} shows the combined spectrum of NGC 4388. The fit, which allowed for flux variations between the various non-simultaneous observations (see Sect.~5), gave a $\chi^2_\nu = 1.41$ for 1431 degrees of freedom.  The normalization factors are in the range 2.5 to 4.2 relative to the {\it XMM-Newton} July 2002 data, which apparently represents the lowest flux level reported so far.   
The power law, dominating the emission above $\sim 2.5 \, \rm keV$, has a photon index of $\Gamma = 1.65 {+0.04 \atop -0.04}$ and an absorption of $N_H = 2.73 {+0.07 \atop -0.07} \times 10^{23} \, \rm cm^{-2}$.

Using a more complicate model with a cut-off power law plus reflection from cold material (the so-called PEXRAV model; Magdziarz \& Zdziarski 1995) instead of the simple power law, does not improve the fit significantly. The data do not constrain the power law and the reflection component: Fixing the photon index to $\Gamma = 1.7$ results in a folding energy of $380 {+\infty \atop -200} \, \rm keV$, indicating that a cut-off is possible but cannot be confirmed.

Applying a single-power law plus an unresolved set of lines, represented by a broad Gaussian line around 1 keV, instead of the Raymond-Smith component does not give acceptable fit results ($\chi^2_\nu \gg 2.0$). 

\begin{figure}
\plotone{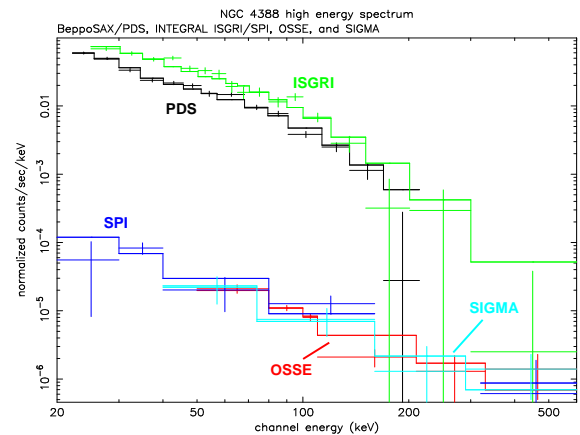}
\caption[]{NGC 4388 combined high energy spectrum, including {\it BeppoSAX} PDS, {\it CGRO} OSSE, {\it SIGMA}, and {\it INTEGRAL} ISGRI/SPI. The spectrum is shown in instrument dependent $\rm counts \, s^{-1} \, keV^{-1}$.}
\label{fig:combinedhigh}
\end{figure}

Using only the high energy ($> 20 \, \rm keV$) data gives a single power law with $\Gamma = 1.72 {+0.05 \atop -0.05}$ with $\chi^2_\nu = 1.5$ (Fig.~\ref{fig:combinedhigh}). Also in this energy region alone the more complex PEXRAV model does not give a significantly better fit. Statistics do not allow us to distinguish between a single power law 
%($2.2 \times 10^{-6} \, \rm ph \, cm^{-2} \, s^{-1} \, keV^{-1}$ and $1.1 \times 10^{-6} \, \rm ph \, cm^{-2} \, cm^{-1} \, keV^{-1}$ at 200 and 300 keV, respectively) 
and a PEXRAV model. 
%($1.7 \times 10^{-6} \, \rm ph \, cm^{-2} \, cm^{-1}\, keV^{-1}$ and $0.6 \times 10^{-6} \, \rm ph \, cm^{-2} \, cm^{-1}\, keV^{-1}$). 
The differences are smaller than the $1\sigma$ error bars of the spectra as seen in Fig.~\ref{fig:combinedhigh} and Fig.~\ref{fig:combined}.

\section{Discussion}

The combination of the {\it INTEGRAL} data with previous observations by {\it XMM-Newton}, {\it BeppoSAX}, {\it CGRO}, and {\it SIGMA} shows that the spectrum from 0.2 keV up to several hundred keV is well represented by a thermal Raymond-Smith model with only Galactic absorption applied, plus a highly absorbed ($N_H \simeq 2.7 \times 10^{23}  \, \rm cm^{-2}$) power law like emission ($\Gamma = 1.7$), and two Gaussian lines to model the iron fluorescence line at $6.39 \, \rm keV$ and $6.9 \, \rm keV$, respectively. The results are consistent with previous studies of NGC 4388 using subsets of these data \citep{SAXNGC,SIGMANGC}. Also observations with {\it Chandra} \cite{CHANDRANGC} and {\it ASCA} \citep{ASCAold,ASCANGC} show a similar spectral behavior. 

The fact that the data of the different missions give an acceptable fit using the same normalization is based on the fact that the error bars in the high energy domain are still rather large. The acceptable correspondence between the ISGRI and SPI data leads to the assumption that the {\it INTEGRAL} calibration is reasonable and that normalization problems as reported e.g. in Courvoisier et al. (2003a) are not as important anymore. But large error bars on the high energy flux measurements (Fig.~\ref{fig:lightcurve}) still leave open the possibility of some errors in the calibration (Lubi\'nski et al. 2004).

The spectral data of NGC 4388 do not drop drastically above 200 keV, but 
are not of high enough significance to distinguish whether there is a cut-off, as expected from Comptonisation models \cite{POP}. On the other hand the absence of a cut-off would be in agreement with Deluit \& Courvoisier (2003), who find a cut-off in the {\it BeppoSAX}/PDS spectra of Seyfert 1, but not in Seyfert 2.  
The {\it XMM-Newton} data show variability only in the energy region above $\sim 2.5 \, \rm keV$. This variability cannot be explained by a change in the absorption column density. But using the same model parameters of the July observation (Tab.~\ref{XMMfit}) for the December 2002 data and allowing the normalization of the single power-law to vary, shows that a flux increase by a factor of $3.4$ is sufficient to model the December data ($\chi^2_\nu = 1.1$; Fig.~\ref{fig:compareMOS}). The spectral shape of the power law, which describes the hard X-rays, the Fe K$\alpha$ line strength, and the soft X-ray component do not seem to vary.

\begin{figure}
\plotone{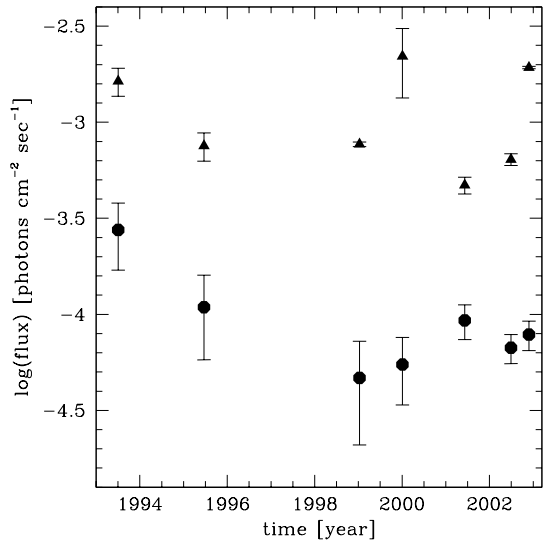}
\caption[]{Flux in the 2-10 keV band (triangles) and of the 6.4 keV Fe K$\alpha$ line (circles) measured by {\it ASCA} (1993 and 1995), {\it BeppoSAX} (1999 and 2000), {\it Chandra} (2001), and {\it XMM-Newton} (2002). The underlying continuum is highly variable, while the line flux shows no significant variation, except for the {\it ASCA} measurement in 1993.}
\label{fig:linevariation}
\end{figure}

The observed spectrum can be compared to the one derived by Ghisellini, Haardt, \& Matt (1994) based on Monte-Carlo simulations for a Seyfert galaxy with inclination $i = 60^\circ - 63^\circ$, column density of $10^{24} \, \rm cm^{-2}$ and a torus with half-opening angle of $\Theta = 30^{\circ}$. 
Based on these assumptions, an equivalent width of the iron fluorescence line of the order of $EW \sim 100 \, \rm eV$ would be expected, which would also mean a parallel evolution of the line flux with respect to the continuum. But 
the Fe K$\alpha$ does not show variability, while the underlying continuum varies by a factor of $\sim 4$ in the line region (see Fig.~\ref{fig:linevariation}). The line flux of $(6.7 \pm 1.2) \times 10^{-5} \, \rm ph \, cm^{-2} \, s^{-1}$ and $(7.8 \pm 1.4) \times 10^{-5} \, \rm ph \, cm^{-2} \, s^{-1}$ for the July and December measurement
% width ($EW = 190 \pm 50 \, \rm eV$) 
is also consistent with the one measured by 
%{\it ASCA} ($EW = 720 \, \rm eV$) and 
{\it Chandra} ($(9.3 \pm 1.9) \times 10^{-5} \, \rm ph \, cm^{-2} \, s^{-1}$) \cite{CHANDRANGC}.
The equivalent width therefore varied strongly ($EW = 190 - 720 \, \rm eV$). 
The exception from this rule is the 1993 ASCA observation, which revealed a high continuum flux with simultaneous high iron line flux.
But the rather stable line flux over several years with simultaneous strong variations of the underlying continuum implies that the line emitting region is separated from the continuum source of the object by several light years.
A similar behavior has been observed in {\it RXTE} spectra of some Seyfert 1 galaxies, where the Fe K$\alpha$ line also varies less strongly than the broadband continuum (Markowitz, Edelson, \& Vaughan 2003).

Another difference of the spectrum observed here to the simulated one by Ghisellini, Haardt, \& Matt (1994) is the fact that the spectrum of NGC 4388 does not seem to have a strong cut-off in the hard X-rays. This could be caused by a higher temperature of the corona ($\gg 100 \, \rm keV$) in the case of NGC 4388. 
If the non-detection of a reflection hump in the hard X-rays is real, this might be a hint for a non isotropic radiation or, generally speaking, a more complex geometry (see e.g. George \& Fabian 1991).
Apparently there is no scattering of the hard X-rays away from the line of sight, except for the iron K$\alpha$ flourescence line. This indicates that in the case of NGC 4388 we indeed observe the unscattered emission of the central engine at hard X-rays. Also no significant decline due to Klein-Nishina processes appears above $50 \, \rm keV$ can be detected in the present data.

Most Seyfert galaxies show a softening of the X-ray continuum as sources brighten (Markowitz, Edelson, \& Vaughan 2003; Nandra et al. 1997).
The fact that the change in luminosity of NGC 4388 in the hard X-rays, and therefore most probably a change in the accretion rate of the central engine, is not accompanied by variations of the spectral shape remains enigmatic. One explanation might be a generally different radiation processing in Seyfert 1 and Seyfert 2, as indicated in e.g. Deluit \& Courvoisier (2003).
Matt et al. (2003) argue that absorption in Seyfert 2 galaxies can be caused by several circumnuclear regions and favour a model with a possible temporary switching-off of the nuclear radiation to explain observed variability.
Their model predicts an iron K$\alpha$ equivalent width of $\sim 100 \, \rm eV$ for a Compton thick case with $N_H \simeq 10^{23} \, \rm cm^{-2}$, which is signifcantly lower than the EW measured in NGC 4388 with a similar absorption column density.
% Matt et al. 2003: Compton thin case, EW 100 eV for NH=1e23
% here: iron line as from a compton thick (>1e24) material
%
% Bianchi et al. 2003: NGC5506, Compton thin (4e22); line variability not significant

The constant soft X-ray emission is most likely not linked to the AGN. This is also supported by the fact, that this emission is extended, as has been reported from {\it ROSAT}/HRI measurements \cite{HRINGC} and lately from {\it Chandra} \cite{CHANDRANGC}. The low abundance in the extended emission has been seen already in the {\it ASCA} observations, though with much larger uncertainties ($Z = 0.05 {+0.35 \atop -0.03} Z_\odot$; Iwasawa et al. 1997). 
Iwasawa et al. (2003) argue that the extended emission at low energies is more likely to origin from photoionized plasma. Even though it is not possible, based on the {\it XMM-Newton} data presented here alone, to give a preference to a thermal or photoionized plasma, the latter one has the advantage of avoiding the low abundance in the vicinity (within several kpc) of the AGN core. 
We therefore refer to the more simple model, applied to the data presented here, which gives sufficient good fit results, even though it might be less meaningful in a physical context.
The {\it Chandra} data also allowed a space resolved study at soft X-rays, which was not possible with {\it XMM}, and found that the AGN core itself apparently contributes less than 10\% to the soft X-ray flux (0.5 -- 2.0 keV) compared to the surrounding gas. 

The {\it INTEGRAL}/ISGRI measurements indicate that a lower flux in the soft gamma-ray region is accompanied by a harder spectrum, but the errors on the spectral slope measurement are too large to be statistically significant. Since the SPI data at the same time show the contrary spectral evolution, it is likely that the spectral slope did not change and that there is only a change in flux ($20 - 200 \rm \, keV$) by a factor of $\sim 1.5$.

This variation is not seen in the {\it RXTE}/ASM data, which cover the softer energy range (1.5 -- 12 keV). Note also the large error values, which are consistent with no flux variation for both, the {\it INTEGRAL} and the {\it RXTE} data. 

\section{Conclusion}

The complex X-ray spectrum of NGC 4388 is composed out of three major components, which originate from different regions:
\begin{itemize}
\item {\bf Hot plasma component}: Emission below $\sim 2.5 \, \rm keV$ is dominated by emission of a hot ($10^7$ $^\circ \rm K$), optically thin plasma with low abundance (about 7 \% solar). This component does not seem to be variable. In {\it ROSAT}/HRI and {\it Chandra} observations this emission appeared to be an extended X-ray nebula out to several kpc from the AGN. {\it Chandra} observations imply that this component might be due to photoionized gas.
\item {\bf Hard X-ray emission}: Emission above $\sim 2.5 \, \rm keV$ follows a highly absorbed ($N_H \sim 2.7 \times 10^{23}  \, \rm cm^{-2}$) power law with photon index $\Gamma = 1.7$ and is related directly to the AGN. While the flux of this component varied by a factor $\simeq 4$ over the past ten years, the spectral shape has been constant within the error bars. No reflection component or cut-off is detectable in the data so far.
\item {\bf Iron line emission}: The Fe K$\alpha$ fluorescence line at $6.39 \rm \, keV$ and K$\beta$ at $6.9 \rm \, keV$ are both consistent with the redshift of the AGN. The former had a constant flux of $\sim 7.5 \times 10^{-5} \, \rm ph \, cm^{-2} \, s^{-1}$ from 1995 to 2002.  The iron line emission is decoupled from the broadband continuum, similar to the behavior observed in several Seyfert 1 galaxies (Markowitz, Edelson, \& Vaughan 2003) and NELG (e.g. NGC 5506; Bianchi et al. 2003). Also the fact that no line broadening is seen supports that it is probably generated several light years away from the AGN core, for example in the molecular torus. 
\end{itemize}

The {\it INTEGRAL} data do not allow to constrain the model on the high energy emission, i.e. there is no sign (except the iron fluorescence line) of a Compton reflection component or a cutoff at higher energies but its existence can also not be ruled out.

Following the observations reported here, NGC 4388 was studied in a dedicated {\it INTEGRAL} observation, which was performed in staring mode in July 2003. Though staring mode observations usually do not produce useful SPI spectra for sources as faint as NGC 4388, the combination of ISGRI and JEM-X data might show in more detail the connection between the hard and soft X-rays, as seen in this work from simultaneous {\it INTEGRAL} and {\it RXTE} data.
Another 100 ksec TOO observation in June 2004 and 500 ksec open time observation of 3C273 in {\it INTEGRAL}'s AO-2 might give further insights into the cut-off of the NGC 4388 spectrum at high energies, e.g. when combined with the already existing data presented here.

\begin{acknowledgements}
This research has made use of the NASA/IPAC Extragalactic Database (NED) which is operated by the Jet Propulsion Laboratory. We thank the anonymous referee for constructive comments, which helped to improve the manuscript.
% and of the SIMBAD Astronomical Database which is operated by the Centre de Donn\'ees astronomiques de Strasbourg.

\end{acknowledgements}

%
% The figures:
%
%
%\clearpage
%
%
% The tables:
%
%\clearpage
\begin{table*}
\caption[]{{\it INTEGRAL} observations}
\begin{tabular}{lcrrc}
\tableline\tableline
obs. date & {\it INTEGRAL}   & SPI             & number of & observation \\
start     & revolution & exp. time [sec] & pointings & mode        \\
\hline
 05/01/03 & 28         &  80520          & 39        & $5 \times 5$\\
 11/01/03 & 30         &  10558          & 2         & staring\\
 17/01/03 & 32         & 182643          & 43        & $5 \times 5$\\
 16/06/03 & 82         &  20000          & 1         & staring\\
 06/07/03 & 89         & 152950          & 47        & $5 \times 5$\\
 09/07/03 & 90         &  59989          & 17        & $5 \times 5$\\
 18/07/03 & 93         &   5283          & 2         & staring\\
% 511943 seconds total ONTIME
\tableline
\end{tabular}
\label{journal}
\end{table*}
\begin{table*}
\caption[]{Fit results for a single-power law model}
\begin{tabular}{lccccc}   
\tableline\tableline
Observation & Photon & $F_X^{a}$ & $F_X^{a}$  &  $F_X^{a}$  & $\chi_{\nu}^2 (dof)$\\
            & Index  & 20-40 keV & 40-100 keV &  100-200 keV &\\  
\hline
Jan.03 / ISGRI  & 1.61 $+0.19 \atop -0.18$ & $4.5 \pm 0.9$ & $8.2 \pm 3.1$ & $8.5 \pm 5.9$ & 1.16 (8)\\
Jan.03 / SPI    & 1.56 $+0.66 \atop -0.53$ & $9.7 \pm 3.6$ & $17.0 \pm 7.8$ & $19.9 \pm 6.2$ & 0.78 (2)\\
Jun.03 / ISGRI  & 1.76 $+0.16 \atop -0.14$ & $6.6 \pm 1.2$ & $10.6 \pm 2.8$ & $9.7  \pm 4.5$& 1.42 (8)\\
Jun.03 / SPI    & 1.29 $+0.37 \atop -0.27$ & $11.0 \pm 3.0$ & $24.3 \pm 8.0$ & $34.0 \pm 11.6$ & 0.70 (5)\\
summed / ISGRI  & 1.70 $+0.01 \atop -0.01$ & $7.0 \pm 0.4$ & $11.7 \pm 1.4$ & $11.3 \pm 3.3$& 1.69 (14)\\
summed / SPI    & 1.68 $+0.47 \atop -0.35$ & $7.7 \pm 2.1$ &$12.2 \pm 4.8$ &  $12.6 \pm 7.7$ & 1.30 (4)\\

\tableline
\end{tabular}
\label{INTEGRALfit}

$^{a}$ un-absorbed flux in $10^{-11}\; \rm erg\,cm^{-2}\,s^{-1}$\\ 
\end{table*}
\begin{table*}
\caption[]{{\it XMM-Newton} observations}
\begin{tabular}{lrrrrr}
\tableline\tableline
obs. date & MOS-1   & MOS-2 & PN\\
start     & exp. time [sec] &exp. time [sec] & exp. time [sec]\\
\hline	  
 07/07/02 &  9850   & 9937  & 4540\\
 12/12/02 &  11667  & 11667 & 8291\\
\tableline
\end{tabular}
\label{journalXMM}
\end{table*}
%
%\begin{table*}
%\caption[]{Journal of {\it XMM-Newton} observations}
%\begin{tabular}{lcrrrr}
%\tableline\tableline
%obs. date & {\it INTEGRAL}   & MOS-1   & MOS-2 & RGS1 & RGS2 \\
%start     & revolution & exp. time [sec] &exp. time [sec] &exp. time [sec] &exp. time [sec]        \\
%\hline
% 07/07/02 & 28         &  9850           & 9937      & 11277 & 10842 \\
% 12/12/02 & 30         &  11667          & 11667     & 11885 & 11885 \\
%\tableline
%\end{tabular}
%\label{journalXMM}
%\end{table*}
%
\begin{table*}
\caption[]{Fit results for the {\it XMM-Newton} EPIC data}
\begin{tabular}{lccccccc}   

Observation & $T [kT]$ & abundance & $N_H$  & $f_{\rm Fe \, K\alpha}$ & $EW$ & $\sigma$ &  $\chi_{\nu}^2 (dof)$ \\
            &          &           & $[10^{23} \rm \, cm^{-2}]$ & $[10^{-5} \rm \, ph \, cm^{-2} \, s^{-1}$]& $\rm \, [eV]$ & $\rm \, [eV]$ & \\
\hline
July 2002 & $0.81 {+0.04 \atop -0.05}$ & $0.08 {+0.03 \atop -0.02}$ & $2.45 {+0.20 \atop -0.21}$ & $6.7 \pm 1.2$ & 566 & $67 \pm 16$ & 1.06 (526)\\ 
Dec. 2002 & $0.80 {+0.03 \atop -0.03}$ & $0.06 {+0.01 \atop -0.01}$ & $2.79 {+0.07 \atop -0.07}$ & $7.8 \pm 1.4$ & 220 & $69 \pm 12$ & 1.25 (861) \\
\end{tabular}
\label{XMMfit}
%$^{a}$ un-absorbed flux in $10^{-11}\; \rm erg\,cm^{-2}\,s^{-1}$\\ 
\end{table*}
\end{document}